\documentclass[amsmath,amssymb,epsfig,twocolumn, showpacs]{revtex4}

\usepackage{graphicx}
\usepackage{dcolumn}
\usepackage{bm}
\begin{document}

\title{Complexes of stationary domain walls in the 
resonantly
forced Ginsburg-Landau equation}
\author{I.V. Barashenkov}
 \email{igor@cenerentola.mth.uct.ac.za}
\author{S.R. Woodford}
 \email{woodford@giulietta.mth.uct.ac.za}
\affiliation{Department of Applied Mathematics,
University of Cape
Town, Rondebosch 7701, South Africa}

\date{\today}
 
\begin{abstract}
The parametrically driven Ginsburg-Landau equation 
has well-known stationary solutions --- the  so-called Bloch and N\'eel, or Ising, walls. In
this paper, we construct an explicit stationary solution describing a bound state of two walls. We also demonstrate that stationary complexes
of more than two walls do not exist. 
\end{abstract}

\pacs{05.45.Yv, 89.75.Kd, 42.65.Tg, 75.60.Ch, 02.30.Ik}

\maketitle

\section{Introduction}
\label{Intro}

In this paper we derive a new class of explicit and physically meaningful solutions to an equation that has been under scrutiny, in various
contexts, for more than $40$\ years. The equation is the resonantly driven Ginsburg-Landau; in its most general form it reads:
\begin{equation}
\label{1}
\psi_t = (\mu + i\nu)\psi  + (g_1+ic_1)\psi_{xx} - (g_3+ic_3)|\psi|^2\psi  - h(\psi^*)^{n-1}.
\end{equation}
Eq.(\ref{1}) describes a one-dimensional chain of coupled weakly nonlinear self-sustained oscillators in the continuum approximation. The
chain is subjected to the periodic forcing at the frequency $\Omega  \approx n\omega_0$, where $\omega_0$\ is the frequency of the undriven
spatially
homogeneous linear oscillations. The complex variable $\psi = \psi(x,t)$\ is the slowly varying amplitude of the resulting nonlinear
oscillations with the period $T = 2\pi n /\Omega$. 

Before specialising Eq.(\ref{1}) to the particular case that will concern us in this paper, we briefly comment on the physical meaning of its
coefficients. First of all, the parameter $\mu > 0$\ measures the distance to the supercritical Hopf bifurcation at which a
(spatially homogeneous) stable 
limit cycle appears. For the stability of the self-sustained oscillation one also needs $g_3 > 0$. The real
constant $\nu$\ is proportional to the frequency detuning, $\nu \propto \Omega/n - \omega_0$. In the derivation of (\ref{1}), it is assumed
that both the linear growth rate $\mu$\ and the detuning $\nu$\ are small compared to the frequency $\omega_0$. (See e.g. \cite{Pikovsky}.)
Next, 
the parameter $c_3$\ describes the nonlinear frequency shift, while 
the derivative term accounts for the interaction of the neighbouring
oscillators  in the chain, with the real and imaginary part of the coefficient $g_1 + ic_1$\ pertaining to the dissipative and reactive
type of coupling, respectively. Finally, $h$\ is proportional to the amplitude of the forcing. The $(n-1)$-st power of $\psi^*$\ in
Eq.(\ref{1}) results from the resonance forcing of order $n:1$\ \cite{Coullet1985}. The $1:1$, $2:1$, $3:1$\ and  $4:1$\ resonances
have been studied most extensively in literature. In this work, we focus on the case $n=2$. 

The analyses of Eq.(\ref{1}) usually start with considering the gradient, or variational, limit \cite{Coullet1990,Malomed_Nepom}. 
The variational limit corresponds to the
assumption that the coupling is purely
dissipative ($c_1 = 0$), frequency detuning $\nu$\ is zero and the homogeneous oscillations are isochronous ($c_3 = 0$). 
With these assumptions, a suitable rescaling of $t$, $x$, $\psi$\ and $h$\ produces 
\begin{equation}
\label{2}
\psi_t = \frac 12\psi_{xx} - |\psi|^2\psi + \psi - h\psi^*.
\end{equation}

In this paper, we consider stationary solutions of Eq.(\ref{2}); these satisfy
\begin{equation}
\label{ODE}
\frac 12 \psi_{xx} - |\psi|^2\psi + \psi - h\psi^{*} = 0.
\end{equation}
We will take $h$\ to be positive in what follows; this can always be achieved by an appropriate phase-shift of  $\psi$. 
Historically, Eq.(\ref{ODE}) was first introduced in the context of the anisotropic $XY$-model, which was used to model an easy-axis
ferromagnet near the Curie temperature \cite{XY,Niez2}. 
Nonstationary magnetization configurations were considered as solutions to Eq.(\ref{2})
\cite{Coullet_Lega_Pomeau}. 
The investigations 
of the more general Eq.(\ref{1}) (still with $n = 2$, though), including analyses of small nonvariational
effects, were reported in \cite{Coullet1990,Skryabin,Park}. Apart from the magnetic 
applications, these studies
were motivated by research in liquid crystals \cite{Nasuno}, experiments with the periodically forced light-sensitive
Belousov-Zhabotinsky reaction \cite{Petrov} and work done in optics, in particular with regard to optical parametric oscillators
\cite{Longhi1} and lasers with intracavity parametric amplification \cite{Longhi2}. Eq.(\ref{1}) also
appeared as a phenomenological equation for the parametrically excited surface waves in viscous fluids \cite{Zhang} and granular
media \cite{Tsimring}.

The nontrivial (spatially nonhomogeneous) solutions which attracted interest in the context of each of these fields are domain walls, or
kinks, also known as dark solitons in nonlinear optics. The domain wall is a localised interface between two different homogeneous
backgrounds (known as domains in the magnetic context). The possible backgrounds are described by the 
constant nonzero
solutions of Eq.(\ref{ODE}), $\psi = \pm A_-$\ and $\psi = \pm iA_+$, where $A_{\pm} = \sqrt{1\pm h}$.  The first of these
is known to be unstable while the second is 
stable. We will only consider dark solitons propagating over
the stable background --- asymptotically, all of the
solutions considered here will satisfy $|\psi|^2 \rightarrow A_+^2$\ as $|x|\rightarrow \infty$.

The soliton solutions to Eq.(\ref{ODE}) with the desired asymptotic behaviour 
can either be topological (with a phase difference of  $180^{\circ}$\ between the two asymptotic values) or nontopological
(with no change in phase between the asymptotic fields).
Eq.(\ref{ODE}) admits two explicit topological solutions. The first is the N\'eel wall \cite{XY,Niez2,Raj,Elphick_Meron},
\begin{equation}
\label{Neel}
\psi_{\rm N} (x) = iA_+ \tanh(A_+ x),
\end{equation}
so named because its magnitude $|\psi|$\ 
vanishes at the origin, at which point the phase becomes discontinuous. 
(In magnetism, a N\'eel wall is a domain
wall with vanishing magnitude  of the magnetisation vector  at its centre.) It is also called the Ising wall, the name appealing to the Ising
model which was used to emulate the easy-axis ferromagnet. The N\'eel wall exists for all positive $h$.

The second topological solution has the form
\begin{equation}
\label{Bloch}
\psi_{\rm B} (x) = iA_+ \tanh(Bx) \pm C\mbox{sech}(Bx),
\end{equation} 
where $B = \sqrt{4h}$\ and $C = \sqrt{1-3h}$\ \cite{Niez2,Sarker,Montonen}.
This solution  is usually referred to as a
Bloch wall (which, in magnetism, is a domain wall  connecting the two domains smoothly,
with the magnetisation vector  remaining nonzero everywhere).   
The Bloch wall  exists in two chiralities, distinguished by the sign of the real part in (\ref{Bloch}).
For convenience, we will refer to the solution with positive
(negative)
real part as the left-handed (right-handed) 
Bloch wall.
Regardless of the  chirality, the Bloch walls only exist  for  $h < \frac 13$. 

\begin{figure}
\includegraphics[ height = 2.6in, width = 1.\linewidth]{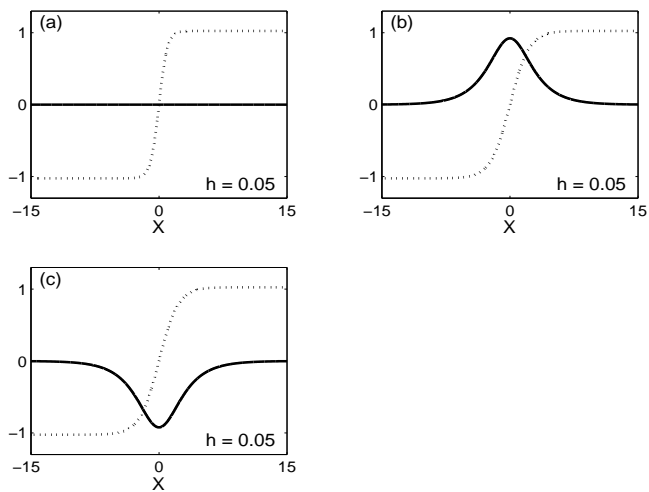}
\caption{\sf The N\'eel wall (a), and the left- and right-handed Bloch walls [(b) and (c) respectively].
The solid line corresponds to the real part, and the dotted line to the imaginary part. Here $h = 0.05$; we plot the two walls for a small
value of the parameter in order to accentuate the difference in their widths.}
\label{Gaylordssometimessayyes}
\end{figure}

In addition to the Bloch and N\'eel walls, Eq.(\ref{ODE}) possesses
nontopological solitons. One such solution
is known explicitly, for $h = \frac{1}{15}$\  \cite{Sarker,SubTrull}:
\begin{equation}
\label{particular}
\psi = iA\left[1 - \frac 32 \mbox{sech}^2(Bx)\right] + 3B\tanh(Bx)\mbox{sech}(Bx).
\end{equation}
Here $B =  \sqrt{4h}=\sqrt{4 / 15}$. In addition, a class of bubble-like solutions (including Eq.(\ref{particular}) as a particular case) 
was found numerically \cite{Ourpaper}. 
As is usual with numerical solutions, the structure of the solitonic bubbles has not been
completely understood. It has also remained unclear
whether they form a continuous family (families) and if they do,  what is their range of existence (in $h$). Finally, a pertinent question is
of the generality of these solutions: can the resonantly driven Ginsburg-Landau equation 
support localised, stationary structures other than the walls and bubbles? The aim of this
paper is to address all of these issues.

Here, for each $h < \frac 13$, we construct a two-parameter family of 
{\it explicit\/} bubble-like solutions and demonstrate that they can
be interpreted as stationary bound states, or complexes, of a Bloch and N\'eel wall. 
Physically, the constructed solutions represent phase domains of
arbitrary length,
which are $180^{\circ}$\ 
out of phase with the background field. We also prove that these solutions exhaust the list of possible stationary complexes --- there
 can be no stationary bound states of more than two walls.

It is appropriate to mention here that equation (\ref{ODE}) appeared previously in several other,  unrelated,  physical 
contexts. Written in terms of the real and imaginary parts of $\psi$, it
coincides with  the
stationary equation for the  so-called MSTB model of field theory
\cite{Montonen,Sarker}. More recently, it occurred as a stationary limit of the parametrically driven 
nonlinear Schr\"odinger \cite{Ourpaper} equation. The latter equation arises in a large variety of physical applications
including nearly resonant optical parametric oscillators \cite{Trillo}, 
easy-plane ferromagnets in magnetic fields \cite{Ourpaper}, and surface waves in wide,
vertically vibrated channels of inviscid fluid \cite{Miles,Elphick_Meron}. Accordingly, we expect the new solutions to admit physical
interpretations in these fields as well.

The paper is organised as follows.  In section \ref{Hirota} we derive a family of explicit two-soliton solutions and cast it in a symmetric
form allowing easy visualisation. These solutions will be interpreted as bound states of a Bloch and N\'eel wall.
 Several families of singular solutions 
appear as by-products in this construction; these will be used later for auxiliary
purposes.  After that (section \ref{CQODE}) we 
 show that there exist no other bounded solutions of Eq.(\ref{ODE}) which would asymptotically approach 
the stable ($\pm iA_+$) background. This means that 
apart from the constant solution, the only nonsingular solutions to
Eq.(\ref{ODE}) are the Bloch and N\'eel walls, and the Bloch-N\'eel complexes constructed in this paper. Finally, 
some concluding remarks are made in section \ref{Conclusions}.

\section{Exact solutions for the Bloch-N\'eel bound state}
\label{Hirota}

\subsection{The Hirota construction}

In order to construct a bound state of two walls explicitly, we employ the Hirota
bilinear formalism (see e.g. \cite{Ablowitz}). Letting
\begin{equation}\label{H2}
  \psi = \frac GF,
\end{equation}
where $G$ is complex and $F$ real function of $x$, equation
(\ref{ODE}) is cast in the bilinear form:
\begin{eqnarray}
F[D^2_x G\cdot F + (2-\lambda) GF - 2hG^*F] \nonumber \\
- G[D^2_x F\cdot F - \lambda F^2 + 2|G|^2] =  0. \label{H3}
\end{eqnarray}
Here $D_x$\ is the Hirota operator defined on ordered products of
functions:
\[
  D_x^n \, a\cdot b \equiv (\partial_x - \partial_y)^n a(x)b(y)|_{x=y}.
\]
In (\ref{H3}) we have added the term $\lambda GF^2$ to the second
bracket and subtracted it from the first one.   The constant
$\lambda$ will be chosen later.

By making the substitution (\ref{H2}) we increased the
number of unknowns while the number of equations remained
unchanged.   We can use this freedom to set the first and
the second term in (\ref{H3}) to zero (separately):
\begin{equation}\label{H4}
  D^2_x \, u \cdot F + (2A^2_- - \lambda) \, uF = 0,
\end{equation}
\begin{equation}\label{H5}
  D^2_x \, v \cdot F + (2A^2_+ - \lambda)vF = 0,
\end{equation}
\begin{equation}\label{H6}
  D^2_x \,  F \cdot F - \lambda F^2 + 2(u^2 + v^2) = 0,
\end{equation}
where we let $G = u + iv$ and decomposed the first bracket in
(\ref{H3}) into its real and imaginary part.   Now we look for a
solution to the system (\ref{H4})-(\ref{H6}) as a series
\[
u = \epsilon u_1 + \epsilon^2 u_2 + \dots, \quad v = v_0 +
\epsilon v_1 + \epsilon^2 v_2 + \dots,  \]  
\[ F = 1 + \epsilon F_1
+ \epsilon^2F_2 + \dots,
\]
where $\epsilon$ is a formal expansion parameter.
An explicit solution will arise if the series truncates
at a finite power of $\epsilon$.

Substituting into (\ref{H4})-(\ref{H6}), the order $\epsilon^0$
gives
\begin{equation}\label{H7}
  \partial^2_x \, v_0 + (2A^2_+ - \lambda)v_0 = 0,
\end{equation}
\begin{equation}\label{H8}
  2v^2_0 - \lambda = 0.
\end{equation}
We now choose $\lambda = 2A^2_+$.   Then, equations (\ref{H7})-(\ref{H8}) 
give $v_0 = A$.   (Here, and in the rest of the paper, $A$ stands for $A_+$.)  Next, at the order $\epsilon^1$
we obtain
\begin{equation}\label{H9}
  (-\partial^2_x + 4h)u_1 = 0,
\end{equation}
\begin{equation}\label{H10}
  \partial^2_x (AF_1 + v_1) = 0,
\end{equation}
\begin{equation}\label{H11}
  (- \partial^2_x + 2A^2)F_1 = 2A v_1.
\end{equation}
Equation (\ref{H9}) yields $u_1 = e^{\theta_1}$, where $\theta_1 =
2h^{1/2} (x - x_1)$ and $x_1$ is an arbitrary
 constant.   From (\ref{H10}) we
infer $v_1 = - AF_1$.   Substituting this into (\ref{H11}) we get
\[
(- \partial^2_x + 4A^2)F_1 = 0,
\]
whence $F_1 = e^{\theta_2}$, where $\theta_2 = 2A (x - x_2)$ and
$x_2$ is another arbitrary constant.

The order $\epsilon^2$ gives three equations:
\begin{equation}\label{H12}
  (- D^2_x + 4h) (u_2 \cdot 1 + u_1 \cdot F_1) = 0,
\end{equation}
\begin{equation}\label{H13}
  D^2_x (v_0 \cdot F_2 + v_1 \cdot F_1 + v_2 \cdot 1) = 0,
\end{equation}
\begin{equation}\label{H14}
  (- D^2_x + 2A^2) (F_1 \cdot F_1 + 2F_2 \cdot 1) = 2 (u^2_1 +
  v^2_1 + 2v_0 v_2).
\end{equation}
Substituting for $u_1$ and $F_1$ in (\ref{H12}) this equation
becomes
\begin{equation}\label{H15}
  (- \partial^2_x + 4h)u_2 = 4A C_- e^{\theta_1 + \theta_2},
\end{equation}
where 
\begin{equation}
C_-=A-2h^{1/2}.
\label{Cminus}
\end{equation}
Ignoring its homogeneous solution, we get
\begin{equation}\label{H16}
  u_2 = - \frac {A - 2\sqrt h}{A + 2\sqrt h} e^{\theta_1 +
  \theta_2}
  \equiv -\frac{C_-}{C_+}e^{\theta_1+\theta_2}.
\end{equation}
Next, substituting $v_0 = A$ and $v_1 = - AF_1$ into (\ref{H13}),
we obtain $v_2 = -AF_2$, after which equation (\ref{H14}) becomes
\[
(- \partial^2_x + 4A^2) F_2 = e^{2\theta_1},
\]
whence, ignoring again the homogeneous solution,
\begin{equation}\label{H17}
  F_2 = \frac 1{4C_+C_-} e^{2\theta_1}.
\end{equation}
The singularity arising for $A^2 = 4h$ can be removed by letting
$x_1 = \infty$ in $\theta_1$.

To the cubic order in $\epsilon$ we get
\begin{equation}\label{H18}
  (-D^2_x + 4h)(u_1 \cdot F_2 + u_2 \cdot F_1 + u_3 \cdot 1) = 0,
\end{equation}
\begin{equation}\label{H19}
  D^2_x (v_0 \cdot F_3 + v_1 \cdot F_2 + v_2 \cdot F_1 + v_3 \cdot
  1) = 0,
\end{equation}
\begin{eqnarray}
  (-D^2_x + 2A^2)(F_3 \cdot 1 + F_2 \cdot F_1)  \nonumber \\  = 2 (u_1 u_2 + v_1
  v_2 + v_0 v_3).\label{H20}
\end{eqnarray}
Substituting for $u_1, u_2, F_1$ and $F_2$, equation (\ref{H18})
becomes
\[
(- \partial^2_x + 4h) u_3 = 0,
\]
whence $u_3 = 0$.   On the other hand, equations (\ref{H19})-(\ref{H20}) 
reduce to the system
\begin{equation}\label{H21}
  \partial^2_x (AF_3 + v_3) = 2A \frac {C_-}{C_+} e^{2\theta_1 +
  \theta_2},
\end{equation}
\begin{equation}\label{H22}
  (- \partial^2_x + 4A^2)(v_3 - AF_3) = 0,
\end{equation}
which gives $v_3 = AF_3$ and
\begin{equation}\label{H23}
  F_3 = \frac {C_-}{4C_+^3} e^{2\theta_1 + \theta_2}.
\end{equation}
Next, the order $\epsilon^4$ yields
\begin{equation}\label{H24}
  (- D^2_x + 4h)(u_4 \cdot 1 + u_2 \cdot F_2 + u_1 \cdot F_3) =
  0,
\end{equation}
\begin{equation}\label{H25}
  D^2_x (v_4 \cdot 1 + v_3 \cdot F_1 + v_2 \cdot F_2 + v_1 \cdot
  F_3 + v_0 \cdot F_4) = 0,
\end{equation}
\begin{eqnarray}
  (-D^2_x + 2A^2)(2F_4 \cdot 1 + 2F_3 \cdot F_1 + F^2_2) = \nonumber \\ 2(u^2_2
  + v^2_2 + 2v_1 v_3 + 2v_0 v_4),\label{H26}
\end{eqnarray}
where we have taken into account that $u_3 = 0$.  Substituting for
all the variables in (\ref{H24}) we obtain
\[
(- \partial^2_x + 4h) u_4=0,
\]
whence $u_4 = 0$.   Equations (\ref{H25})-(\ref{H26}) reduce to
the homogeneous system
\begin{eqnarray*}
  \partial^2_x (v_4 + AF_4) = 0, \\
(- \partial^2_x + 2A^2)F_4 = 2A v_4.
  \end{eqnarray*}
We choose the trivial solution, $v_4 = F_4 = 0$.

The order $\epsilon^5$ is the last order that we have to do ``by
hand''; in dealing with all higher orders  ($\epsilon^n$ with $n
\ge 6$) we will simply invoke the machinery of mathematical
induction.   To $\epsilon^5$, we get
\begin{equation}\label{H27}
  (-D^2_x + 4h) (u_2 \cdot F_3 + u_5 \cdot 1) = 0 ,
  \end{equation}
\begin{equation}\label{H28}
  D^2_x (v_0 \cdot F_5 + v_2 \cdot F_3 + v_3 \cdot F_2 + v_5 \cdot
  1) = 0,
\end{equation}
\begin{equation}\label{H29}
  (- D^2_x + 2A^2) (F_5 \cdot 1 + F_2 \cdot F_3) = 2(A v_5 + v_2
  v_3),
\end{equation}
where we have taken into account that $u_3 = u_4 = v_4 = F_4 = 0$.
Substituting for $u_2$ and $F_3$, equation (\ref{H27}) gives
\[
(-\partial^2_x + 4h) u_5 = 0,
\]
whence $u_5 = 0$.  Recalling that $v_2 = - A F_2$ and $v_3 = A
F_3$, equation (\ref{H28}) becomes
\[
\partial^2_x (v_5 + A F_5) = 0,
\]
whereas making use of $(-D^2_x + 2A^2) F_2 \cdot F_3 = 2v_2 v_3$ in
(\ref{H29}) we get
\[
(- \partial^2_x + 2A^2) F_5 - 2A v_5 = 0.
\]
The last two equations are satisfied by letting $v_5 = F_5 = 0$.

Finally, we prove that all coefficients $u_n, v_n$, and  $F_n$ with $n
\ge 6$ are also equal to zero.   We assume that $u_{n-1} = v_{n-1}
= F_{n-1} = 0$ and show that this 
entails $u_{n} = v_{n}= F_{n} = 0$.  
Consider, first, equations (\ref{H5})-(\ref{H6}).  
Setting to zero the coefficients of $\epsilon^n$, we
obtain
\begin{equation}\label{H30}
  D^2_x \, \left(v_0 \cdot F_n+\sum^{n-1}_{k=1} v_k \cdot F_{n-k} + v_n \cdot 1
  \right) = 0,
\end{equation}
\begin{eqnarray}
  (-D^2_x + 2A^2)
  \left(\sum^{n-1}_{k=1} F_k \cdot F_{n-k} + 2F_n \cdot
  1\right) \nonumber \\ = 2 \sum^{n-1}_{k=1} (u_k u_{n-k} + v_k v_{n-k}) + 4v_0 v_n.\label{H31}
\end{eqnarray}
Since $u_{n-k} = 0$ for all $3 \le n-k \le n-1$ and $F_{n-k} =
v_{n-k} = 0 $ for all $4 \le n-k \le n-1$, the sum involving
$u_{n-k}$ in the right-hand side of (\ref{H31})
begins with $k = n - 2$ (rather than with $k = 1$),
while all
sums involving  $F_{n-k}$ and $v_{n-k}$ begin with $k = n-3$.  On
the other hand, since $n - 2 \ge 4$, all $u_k$ in the sum in
(\ref{H31}) are equal to zero.   In a similar way, all $v_k$ and
$F_k$ in the sums in (\ref{H30}) and (\ref{H31}) equal zero --
except $v_{n-3}$ and $F_{n-3}$ for $n=6$.   Therefore, for $n \ge
7$ equations (\ref{H30})-(\ref{H31}) become a pair of
homogeneous equations for $v_n$ and $F_n$; hence 
we can set $v_n = F_n = 0$.
For $n = 6$, we get
\begin{eqnarray*}
  D^2_x (v_0 \cdot F_6 + v_3 \cdot F_3 + v_6 \cdot 1)  =  0, \\
  (-D^2_x + 2A^2) (F_3 \cdot F_3 + 2F_6 \cdot 1)  =  2(v_3^2 +
  2v_0 v_6).
\end{eqnarray*}
Using $v_3 = AF_3$, this also reduces to a homogeneous system for
$v_6$ and  $F_6$.

Lastly, equation (\ref{H4}) gives, to the order $\epsilon^n$:
\[
(- D^2_x + 4h) \left(\sum^{n-1}_{k=1} u_k \cdot F_{n-k} + u_n \cdot 1
\right)
= 0.
\]
Since $n - 3 \ge
3$,  we have $u_k = 0$ for $k \ge n - 3$.
On the other hand, all $F_{n-k} = 0$ for $k \le n - 3$
and so all terms in the sum equal zero.
Hence $u_n = 0$. Q.E.D.

\subsection{The explicit solution and its interpretation}

Thus we have constructed an explicit solution of the form $\psi =
(u + iv)F^{-1}$, where $u,v$ and $F$ are polynomials of
$e^{\theta_1}$ and $e^{\theta_2}$ with real coefficients. It will be shown later that if $C_- < 0$, the solution has a singularity. [We
remind that $C_-$\ is given by Eq.(\ref{Cminus}).] For now,
we assume that $C_- > 0$\ and  cast the solution
in a  more symmetric form.

First of all, the parameter $\epsilon$ can be absorbed into
$e^{\theta_1}$ and $e^{\theta_2}$ through the re-definition of the
arbitrary constants $x_1$ and $x_2$.   Next, we define $\chi_1 =
\theta_1 - \alpha - \beta$ and $\chi_2 = \theta_2 - 2\beta$,
where 
\begin{eqnarray}
e^{2\alpha} = 4C_+ C_-=4(A^2-4h), \nonumber \\  e^{2\beta} = \frac{C_+}{C_-}=
\frac{A+2h^{1/2}}{A-2h^{1/2}}.
\label{H370}
\end{eqnarray}
The new phases $\chi_1$ and $\chi_2$ still involve arbitrary
constants $x_1$ and $x_2$.   We can choose these constants in such a way
that, up to an overall translation,
\begin{subequations}
\begin{equation}\label{H35}
  \chi_1 = 2h^{1/2} (x - s),  \quad  \chi_2 = 2A (x + s),
\end{equation}
where $s$ is the only free parameter remaining.
The numerator and denominator of the solution 
\begin{equation}
\psi = (u +
iv)F^{-1}
\label{H36qqqq}
\end{equation} 
are then given by the following expressions:
\label{H36}
\begin{eqnarray}
u = e^{\alpha + \beta + \chi_1} (1 - e^{\chi_2}), \label{H36a} \\
v = A(1 - e^{2\beta + \chi_2} - e^{2\beta + 2\chi_1} + e^{2
\chi_1 + \chi_2}), \label{H36b} \\
F = 1 + e^{2\beta + \chi_2} + e^{2\beta + 2\chi_1} + e^{2\chi_1 + \chi_2}.\label{H36c} 
\end{eqnarray}
\end{subequations}
Asymptotically, $\psi \rightarrow iA$ both  as $x \rightarrow \infty$
and $x \rightarrow -\infty$, and hence Eq.(\ref{H36}) has the form of a bubble.
 If we perform the reflection $x \to - x$, and at the same time
replace $s$ with $-s$, the real and imaginary parts of the solution
change according to $u / F \to - u/F$, $v/F \to 
v/F$. 
 Denoting  the 
solution (\ref{H36}) by $\psi(x;s)$, we therefore have the following symmetry:
\[
\psi(-x;-s) = - \psi^*(x;s).
\]
This implies, in particular, that the solution 
(\ref{H36}) with $s = 0$ has an odd real and even
imaginary part. When $h=\frac{1}{15}$, the $s=0$ solution reproduces the explicit solution that has been known before,
Eq.(\ref{particular}).
(More precisely, it is equivalent to (\ref{particular}) with $x \to -x$.)

The solution (\ref{H36}) describes a stationary complex of a Bloch and a N\'eel wall, with the parameter $s$\ characterising their
separation. This is easily seen by examining Eq.(\ref{H36}) in the limit of large $s$. 
First, let $s$\ be large and positive. 
For  $x \sim s$, we have $e^{\chi_2} \gg e^{2\chi_1} \sim 1$. In this region, the solution (\ref{H36}) reduces to
\[ \psi=iA\tanh X - \sqrt{1-3h} \, \mbox{sech}X, \]
with $X = 2\sqrt{h}(x-s)-\beta$, which is  a right-handed Bloch wall centered at $x_0 = s + \frac 12 \beta h^{-1/2}$. 
In the region $x \sim -s$, we find that (\ref{H36}) 
becomes
\[ \psi=-iA\tanh[A(x+s)+\beta], \]
which is a N\'eel wall centered at $x_0 = -s - \beta A^{-1}$. 
On the other hand, if  we let $s$\ be large and negative, then we find a N\'eel wall
on the right (centered at $x_0 = -s + \beta A^{-1}$) and a
 right-handed Bloch wall on the left (centered at $x_0 = s - \frac 12 \beta h^{-1/2} $). 

\begin{figure}
\includegraphics[ height = 2.6in, width = 1.\linewidth]{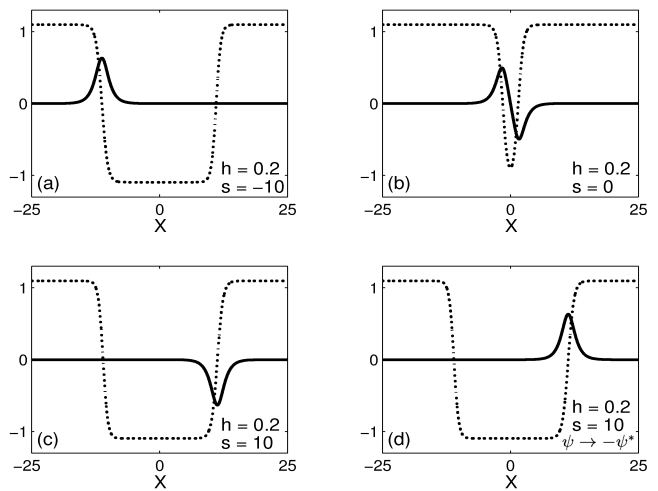}
\caption{\sf The bubble solution (\ref{H36}) for (a) $s = -10$, (b) $s = 0$\ and (c) $s = 10$. The panel (d) corresponds to the bubble of
opposite chirality [i.e. with $\psi(x) \rightarrow -\psi^{*}(x)$]; like (c), it is plotted for $s = 10$.
 The solid line corresponds to the real part, and the dotted line to the imaginary part.}
\label{Gaylordssayno}
\end{figure}

These conclusions are illustrated by Fig.\ref{Gaylordssayno} which depicts the real and imaginary part of (\ref{H36}) for several
representative values of $s$. We also note that the transformation $x\rightarrow -x$, $s \rightarrow -s$\ which was noted above to change the
sign of the real part while leaving the imaginary part intact, simply flips the chirality of the Bloch wall. 
(See Fig.\ref{Gaylordssayno}(d)).

There is yet another way of seeing that Eq.(\ref{H36}) represents a bound state of two walls, and this time we do not have to assume that
$s$\ is large. Consider the integral
\begin{subequations}
\label{Power}
\begin{equation}
I = \int \left( A^2 - |\psi|^2 \right) \, dx
\label{39a}
\end{equation}
which gives an integral measure (``area'') 
of solutions with $|\psi(x)|\to A$\ as $|x|\to \infty$. This integral usually has some physical meaning in
the nonlinear Schr\"odinger-based interpretations of Eq.(\ref{ODE}). For example, if Eq.(\ref{ODE}) is regarded as a stationary reduction of
the Landau-Lifshitz equation for the easy-plane ferromagnet in the external magnetic field \cite{Ourpaper}, the integral (\ref{39a}) gives
the total number of magnons in the excited state of the ferromagnet. In the Ginsburg-Landau based applications of Eq.(\ref{ODE}),
the integral
(\ref{39a}) does not usually have any special meaning, but nevertheless can be used as a scalar characteristic of stationary 
solutions. Letting $\psi =
(u + iv) F^{-1}$, Eq.(\ref{39a}) becomes
\begin{equation}
\label{39b}
I = \int  \left(A^2 F^2 - u^2 -v^2 \right)\, \frac{dx}{F^2}.
\end{equation}
\end{subequations}
For the Bloch wall, this integral equals $I_{\rm B}=2B$ while  for the N\'eel wall we get
$I_{\rm N}=2A$.  
 As for the solutions (\ref{H36}), 
the evaluation of the integral (\ref{Power})  becomes trivial if we notice that, due to Eq.(\ref{H6}), 
the integrand  is a total derivative:
\begin{equation}
 \left(A^2 F^2 - u^2 -v^2 \right) F^{-2} = \partial_x \left(\partial_x F/ F \right).
\end{equation}
Evaluating $\partial_xF/F$ at $x= \pm \infty$ we conclude that
 each member of the family of bubbles (\ref{H36})
has the ``area'' $I=2(A+B)$ --- which is exactly the sum of $I_{\rm B}$ and $I_{\rm N}$.

We conclude this section by mentioning that bound states of a pair of stationary 
dark solitons have been also found in the (integrable) Manakov system \cite{Kivshar2}.

\subsection{Other solutions}

If we had chosen $F_1 = -e^{\theta_2}$\ instead of $e^{\theta_2}$\ at the order $\epsilon$, we would have arrived at a different solution.
Assuming $h < \frac 13$\ (so that $C_- > 0$) and defining $\alpha,\,\beta,\, \chi_1$\ and $\chi_2$\ as in  
Eqs.(\ref{H370}) and (\ref{H35}), this new solution can be written as Eq.(\ref{H36qqqq}) with 
\begin{subequations}
\label{H40}
\begin{eqnarray}
u = e^{\alpha + \beta + \chi_1} (1 + e^{\chi_2}), \label{H40a} \\
v = A(1 + e^{2\beta + \chi_2} - e^{2\beta + 2\chi_1} - e^{2\chi_1 + \chi_2}), \label{H40b} \\
F = 1 - e^{2\beta + \chi_2} + e^{2\beta + 2\chi_1} - e^{2\chi_1 + \chi_2}. \label{H40c}
\end{eqnarray}
\end{subequations}
It is not difficult to check that this solution is singular. Indeed, the function 
$F(x)$\ is continuous, and has opposite signs at the two infinities: 
$F\rightarrow 1$\ as $x\rightarrow -\infty$ and $F\rightarrow -\infty$\ as $x\rightarrow \infty$.
Therefore, it must pass through zero at least once. Since $u$\ cannot vanish for finite $x$, we have $ u/F = \infty$\ 
wherever $F = 0$. Although singular solutions are not physically meaningful, there will be some indirect use for them in the next section.

Assume now $h > \frac 13$. Here  we have $C_- = A - 2h^{1/2}< 0$\ and the functions (\ref{H36}) and (\ref{H40}) are no longer
solutions,
because the parameters $\alpha$\ and $\beta$, as defined by Eq.(\ref{H370}), are not real. We can, however, define
$\alpha$\ and $\beta$\ by
\begin{eqnarray}
e^{2\alpha} = -4C_+ C_-=4(4h - A^2), \nonumber \\  e^{2\beta} = -\frac{C_+}{C_-}=
\frac{2h^{1/2}+A}{2h^{1/2}-A},
\label{H41}
\end{eqnarray}
and, instead of Eqs.(\ref{H36a})-(\ref{H36c}) and (\ref{H40a})-(\ref{H40c}), arrive at the following two solutions:
\begin{subequations}
\label{H42}
\begin{eqnarray}
u = e^{\alpha + \beta + \chi_1} (1 \pm e^{\chi_2}), \label{H42a} \\
 v = A(1 \mp e^{2\beta + \chi_2} + e^{2\beta + 2\chi_1} \mp e^{2
\chi_1 + \chi_2}), \label{H42b} \\
F = 1 \pm e^{2\beta + \chi_2} - e^{2\beta + 2\chi_1} \mp e^{2\chi_1 + \chi_2}. \label{H42c}
\end{eqnarray}
\end{subequations}
If the top sign is chosen in (\ref{H42}), an argument similar to the one following Eq.(\ref{H40}) 
demonstrates that $F$\ must pass through zero whereas $u$\ is strictly positive; once
again, $u/F$\ is singular. If the bottom sign is chosen, $F$\ is positive in the asymptotic regions $x\rightarrow \pm \infty$,
and this type of argument would not work.
However, at the point $x = -s-\beta A^{-1}$, we have  $\chi_2 = -2\beta$\ and so Eq.(\ref{H42c}) (with the bottom sign chosen) becomes 
\begin{equation}
F(x) =e^{2\chi_1}\left(e^{-2\beta} - e^{2\beta}\right). \label{H43}
\end{equation}
By Eq.(\ref{H41}), $e^{2\beta}>1$, which means that the contents of the brackets in 
(\ref{H43}) is negative. Thus for this value of $x$, $F < 0$. But since $F$\ is positive in both asymptotic regions and is continuous,
it follows that $F$\ must
vanish at least at two points.
Since the bottom sign has been chosen, (\ref{H42b}) implies that $v \geq A > 0$, so when $F = 0$, the quotient $v/F$\
is infinite.

Thus, in summary, of all the families of solutions found in this section, only  family 
(\ref{H36}) is regular, and only for $h<\frac 13$. The other
solutions are singular and not physically relevant.

\section{``Completeness'' of the list of solutions}
\label{CQODE}

It is convenient to rescale Eq.(\ref{ODE}) so that the asymptotic values are equal to $\pm 1$. Letting
$\psi(x) = iA\Psi (Ax)$,  Eq.(\ref{ODE}) becomes 
\begin{equation}
\label{ODE2}
\Psi_{xx} + \frac{2}{A^2}\Psi - 2|\Psi|^2\Psi + 2\left(1 -\frac{1}{A^2} \right)\Psi^* = 0.
\end{equation}

We will demonstrate that solutions (\ref{Neel}), (\ref{Bloch}) and (\ref{H36}) (when suitably rescaled) are the {\it only}
bounded  
solutions of Eq.(\ref{ODE2}) that asymptotically approach one of the stable flat
backgrounds $\Psi = \pm 1$\ as $x\rightarrow -\infty$. We will consider 
solutions of Eq.(\ref{ODE2})
as trajectories in a $4$-dimensional phase space and show that for any direction in which
a trajectory
can leave the fixed point (representing the flat background), we already have a solution leaving in that direction.
By uniqueness,
no other trajectories will be allowed to 
exist. For definiteness, we confine ourselves to the case of trajectories that approach $\Psi = 1$\ as $x\rightarrow -\infty$; a similar
argument is valid for solutions approaching $\Psi = -1$.

For convenience of presentation, we start by listing 
all known solutions to the rescaled equation (\ref{ODE2}). Besides the solutions derived in the previous section, we
also include solutions which are the 
unbounded counterparts of the isolated Bloch and N\'eel walls. In all of these,
we make explicit the translational invariance of Eq.(\ref{ODE2}).

For $h < \frac 13$, apart from the flat background
$\Psi_0 = 1$, the list consists of the N\'eel wall $\Psi_{\rm N} = -\tanh(x-x_0)$\ and its singular counterpart 
\begin{equation}
\label{Neel2}
\tilde\Psi_{\rm N} = -\coth(x-x_0); 
\end{equation}
the Bloch walls
\begin{equation}
\label{Bloch2}
\Psi_{\rm B} = -\tanh \left[B(x-x_0)\right] \pm  i C\mbox{sech}\left[B(x-x_0)\right], 
\end{equation}
where $B = 2(A^2 - 1)^{1/2} A^{-1}$\ and $C = (4-3A^2)^{1/2}A^{-1}$, and, finally, the Bloch-N\'eel complex
(the bubble)  which we write together with its
unbounded counterpart:
\begin{subequations}
\begin{equation}
\label{Bubble2}
\Psi_{\rm BN} = \frac{u + i\sigma v}{1 \pm e^{ 2\beta + \chi_2} + e^{2\beta + 2\chi_1} \pm e^{2\chi_1 + \chi_2}}.
\end{equation}
Here
\label{Whynot?}
\begin{eqnarray}
u = 1 \mp e^{2\beta +\chi_2} - e^{2\beta + 2\chi_1} \pm e^{2\chi_1 + \chi_2}, \label{Whynot1?} \\
v = 2(1+B)e^{\chi_1}(1 \mp e^{\chi_2}), \label{Whynot2?}
\end{eqnarray}
\end{subequations}
$e^{2\beta} = (1+B)(1-B)^{-1}$, $\chi_1 = B(x-x_0-s)$\ and $\chi_2 = 2(x-x_0+s)$. The real parameters $s$\ and $x_0$\ are
arbitrary; $s$\ characterises the separation distance and $x_0$\ describes
uniform translations. The sign factor $\sigma = \pm 1$\ 
determines the chirality of the Bloch wall bound in the complex. [Note that $\sigma$\ is not correlated with the sign factor in
(\ref{Whynot1?})-(\ref{Whynot2?}).] If the top sign 
 is chosen, Eq.(\ref{Whynot?}) gives the rescaled regular solution (\ref{H36}). 
If the bottom sign is chosen, we have the rescaled 
singular  solution (\ref{H40}). 

For $h > \frac 13$, the functions 
$\Psi_0$, $\Psi_{\rm N}$\ and $\tilde\Psi_{\rm N}$\ defined in the previous paragraph 
remain as solutions. Eqs.(\ref{Bloch2}) and (\ref{Whynot?}), on the other hand, 
are no longer solutions.
Eq.(\ref{Bloch2}) is replaced by
\begin{equation}
\label{labelthis}
\tilde\Psi_{\rm B} = -\coth \left[B(x-x_0)\right] \pm  i C\mbox{cosech}\left[B(x-x_0)\right], 
\end{equation}
where $B$\ stays as it was previously defined whereas $C$\ is now given by $C = (3A^2 - 4)^{1/2}A^{-1}$. Eq.(\ref{Whynot?}) is replaced by
\begin{subequations}
\label{goodreason}
\begin{equation}
\tilde\Psi_{\rm BN} = \frac{u + i\sigma v}{ 1 \mp e^{2\beta + \chi_2} - e^{2\beta + 2\chi_1} \pm e^{2\chi_1 + \chi_2}}, 
\end{equation}
where
\begin{eqnarray}
u = 1 \pm e^{2\beta + \chi_2 } + e^{2\beta 2\chi_1 } \pm e^{2\chi_1 + \chi_2}, \label{Whynot1a?} \\
v = 2(1+B)e^{\chi_1}(1 \mp e^{\chi_2}), \label{Whynot2a?} 
\end{eqnarray}
\end{subequations}
and all parameters are defined as for  (\ref{Whynot?}) except  $e^{2\beta} = (B+1)(B-1)^{-1}$.
Both (\ref{labelthis}) and (\ref{goodreason}) are singular solutions [for all choices of signs in (\ref{goodreason})].

We now turn to Eq.(\ref{ODE2}) and write it as a 
dynamical system for a particle on the plane:
\begin{subequations}
\label{system}
\begin{eqnarray}
\eta_{xx} - 2(\eta^2 + \xi^2)\eta + 2\eta = 0, \label{system2.5} 
\\  \xi_{xx} - 2(\eta^2 + \xi^2)\xi + (2-B^2)\xi = 0,  \label{system2}
\end{eqnarray}
\end{subequations}
\noindent where $\Psi = \eta + i\xi$\ with  $\eta$\ and $\xi$\ real, and $B^2 = 4(A^2-1)A^{-2}$\ (as defined for 
Eq.(\ref{Bloch2})). 
This is a hamiltonian system, with the hamiltonian
\begin{equation}
{\cal H} = \frac 12 \left[\eta_x^2+\xi_x^2 - (\eta^2+\xi^2-1)^2  - B^2 \xi^2\right].
\label{sec2hamilt2}
\end{equation}
\noindent There is also a second, independent, integral of motion:
\begin{equation}
{\cal I} = (\xi \eta_x - \eta \xi_x)^2 + B^2\left[\eta_x^2 - \eta^2\xi^2 -(\eta^2-1)^2 \right].
\label{sec2otherint2}
\end{equation}
(In the derivation of (\ref{sec2otherint2}) we were guided by the results of \cite{Hietarinta} which considers a similar system.)
Thus all trajectories of the system (\ref{system}) are confined
 to lie on a
2-dimensional surface, defined by the constraints (\ref{sec2hamilt2}) and (\ref{sec2otherint2}). 

We consider trajectories that flow out of
the fixed point $(\eta,\xi,\eta_x,\xi_x) = (1,0,0,0)$. (On these trajectories, the conserved quantities obey ${\cal H}=0={\cal I}$). 
This
fixed point is a saddle, with two positive and two negative real eigenvalues. Thus in a neighbourhood of the fixed point on the unstable
manifold, $\eta_x$\ will be either positive or negative (as will $\xi_x$), while oscillatory behaviour is not possible. 
In this 
neighbourhood, the sign of $\eta_x$\ will obviously be the same as the sign of $(\eta-1)$; similarly, the sign of $\xi_x$\ will
be the same as that of $(\xi-0)$.
As for the {\it magnitudes} of $\eta_x$\ and $\xi_x$, these are determined by $\eta$\ and $\xi$\ via the  constraints ${\cal H} =
0$, ${\cal I} = 0$. Thus the variables $\eta$\ and $\xi$\ uniquely determine $\eta_x$\ and $\xi_x$\ 
in the vicinity of the fixed point and therefore provide
coordinates on the local unstable manifold.

In the vicinity of  the fixed point, Eqs.(\ref{system2.5}) and (\ref{system2}) can be written as 
\begin{subequations}
\begin{eqnarray}
\eta_{xx} = 4(\eta-1) + 2\xi^2, \\
\xi_{xx} = B^2 \xi.
\end{eqnarray}
\label{equations}
\end{subequations}
(Note that we cannot drop the $\xi^2$\ term in the top equation here, as the equation does not include any term linear in $\xi$.) 
The solution to (\ref{equations}) satisfying $\eta\rightarrow 1,\, \xi\rightarrow 0$\ as $x\rightarrow -\infty$,  is
\begin{subequations}
\label{somethingslightlydifferent}
\begin{equation}
\eta = 1 + {\cal M} e^{2x} - \frac{{\cal N}^2}{2(1-B^2)} e^{2Bx}, \label{something}
\end{equation}
\begin{equation}
\xi = {\cal N} e^{Bx}. \label{something2}
\end{equation}
\end{subequations}
The real 
constants ${\cal M}$\ and ${\cal N}$\ are arbitrary and provide a parametrisation of the local unstable manifold: each pair $({\cal M}, \,
{\cal N})$\ defines a trajectory on the manifold, and the other way around --- 
for each point ($\eta,\,\xi,\,\eta_x,\,\xi_x $) on the manifold, we can find a pair $({\cal M}, \, {\cal N})$\ 
such that there is a  trajectory connecting ($\eta,\,\xi,\,\eta_x,\,\xi_x $) to ($1,\,0,\,0,\,0$) which is 
described by (\ref{somethingslightlydifferent}). 
Indeed, for the given pair of coordinates $(\eta_0,\,\xi_0)$\ we can solve (\ref{somethingslightlydifferent}) to get 
\begin{equation}
\label{starryeyedsurprise}
{\cal M}e^{2x_0} = \eta_0 + \frac{\xi_0^2}{2(1-B^2)} -1, \quad {\cal N}e^{Bx_0} = \xi_0.
\end{equation}
Due to the translational invariance of the system (\ref{system}), replacing $x$\ with $x-x_0$\ in 
(\ref{somethingslightlydifferent}) simply
furnishes a different parametrisation of the same trajectory. Hence the required pair  $({\cal M}, \,{\cal N})$\  is obtained, e.g. 
by setting
$x_0 = 0$\ in (\ref{starryeyedsurprise}).

Thus, if we factor the translation invariance out, there is a one-to-one correspondence between $({\cal M}, \,
{\cal N})$\ and  $(\eta, \, \xi)$\ on the local unstable manifold.
We now show that for
every pair
$({\cal M}, \,{\cal N})$\ (with $-\infty < {\cal M, \, N} < \infty$) we already 
have an explicit solution in our list, regular or singular, with the asymptotics (\ref{somethingslightlydifferent}).
This will imply that our list is complete and no other solutions with this asymptotic behaviour can
exist.

First of all, the solution corresponding to $({\cal M}, \,{\cal N}) = (0,\, 0)$\ is the flat background $\Psi_0 = 1$.
Keeping ${\cal N} = 0$, 
we
have two possibilities (for all $h$):
in 
the case ${\cal M} > 0$, Eqs.(\ref{somethingslightlydifferent})
give the asymptotics of the solution $\tilde{\Psi}_{\rm N}$, Eq.(\ref{Neel2}), with $x_0$\ defined by ${\cal M} = 2e^{-2x_0}$;
similarly, the case  ${\cal M} < 0$\ corresponds to the N\'eel wall $\Psi_{\rm N}$, with ${\cal M} = -2e^{-2x_0}$. 

Now let ${\cal M} = 0$\ while ${\cal N}\neq 0$. Here the two cases, $h<\frac 13$\ and $h > \frac 13$,
have to be considered separately. For $h<\frac 13$,
the pair $(0,\,{\cal N})$\ corresponds to the Bloch wall (\ref{Bloch2}), with 
\begin{equation}
\label{24601}
|{\cal N}| = 2(1-3h)^{1/2}A^{-1}e^{-Bx_0}.
\end{equation}
The sign of ${\cal N}$\ is
arbitrary and determines the chirality of the soliton. If $h>\frac 13$, we recover the solution $\tilde \Psi_{\rm B}$; 
this solution also occurs with two different chiralities, depending on the sign of   ${\cal N}$.

Finally, we let ${\cal M} \neq 0$, ${\cal N} \neq 0$. Assume, first, that  $h <  \frac 13$\ and consider the solution $\Psi_{\rm BN}$, 
Eq.(\ref{Whynot?}). This solution comes in two chiralities, $\sigma = \pm 1$. We let ${\cal N} > 0$; this selects one of
the chiralities ($\sigma = +1$). (The case ${\cal N}< 0$\ can be considered in a similar way.) Comparing the asymptotic behaviour of the
solution (\ref{Whynot?}) to Eqs.(\ref{somethingslightlydifferent}), we find 
\[ {\cal M} = \mp 2 e^{2(\beta - x_0 + s)}\] 
and 
\[{\cal N} =  2 (1+B) e^{-B(x_0 + s)}.\]
Thus ${\cal M} <0$\ corresponds to the (regular) Bloch-N\'eel bound state [top sign in 
(\ref{Whynot?})], and ${\cal M} >0$\ to its singular counterpart [bottom sign in (\ref{Whynot?})]. For $h > \frac 13$, a similar
consideration involving Eq.(\ref{goodreason}) demonstrates that any pair $({\cal M}, \,{\cal N})$\ with 
${\cal M} \neq 0$, ${\cal N} \neq 0$\ corresponds to a known
solution as well. Q.E.D.

\section{Concluding remarks}
\label{Conclusions}

In this paper, we 
have  
explicitly constructed  bound states of a pair of domain walls in the resonantly forced Ginsburg-Landau equation. 
The constructed solutions represent domains (of arbitrary length) that are 
$180^{\circ}$ out of phase with  the background field.  We have 
also demonstrated that stationary bound states of more than two dark solitons cannot
exist. 
(We have, in fact, proved that the solutions presented in this paper are the {\it only\/} solutions to the reduced scalar equation 
(\ref{ODE})
which approach the stable background at infinity.)

Apart from the Ginsburg-Landau equation, Eq.(\ref{ODE}) occurred as a stationary reduction of several conservative nonlinear evolution
equations modelling some other 
physical situations. We have already mentioned the MSTB (Montonen-Sarker-Trullinger-Bishop) model of field
theory \cite{Sarker,Montonen}; the corresponding equation of motion can be written as the complex $\phi^4$-equation with the broken $U(1)$\
symmetry:
\begin{equation}
\frac 12 \psi_{tt} - \frac 12 \psi_{xx} + |\psi|^2\psi - \psi + h\psi^{*} = 0. 
\label{KG}
\end{equation}
(For the current status of the MSTB and related theories, see \cite{Izquierdo}.) In addition, 
Eq.(\ref{ODE}) describes stationary solutions of the
parametrically driven nonlinear Schr\"odinger equation:
\begin{equation} 
i\psi_t + \frac 12 \psi_{xx} - |\psi|^2\psi + \psi - h\psi^{*} = 0. 
\label{NLS}
\end{equation}
In fluid dynamics, (\ref{NLS})
governs the amplitude of
the oscillation of the water surface in a vertically vibrated channel with large width-to-depth 
ratio \cite{Miles,Elphick_Meron}. 
(If, conversely, the channel is deep and narrow, Eq.(\ref{NLS}) is still valid, but with the opposite sign in front of the nonlinear term.)
The same Eq.(\ref{NLS}) arises as an
amplitude
equation for the upper cutoff mode in the parametrically driven 
nonlinear lattices \cite{lattices}.
It was also derived for the
doubly resonant $\chi^{(2)}$
optical parametric oscillator in the limit
of large second-harmonic detuning \cite{Trillo}. 
In all of these cases, the term $h\psi^*$\ represents parametric pumping of some sort. 
Finally, Eq.(\ref{NLS}) describes magnetisation waves in a quasi-one-dimensional ferromagnet with a weakly anisotropic easy plane, in 
a perpendicular stationary magnetic field \cite{Ourpaper}.
In the magnetic context, the $h\psi^*$\ term accounts for  the anisotropy of the ferromagnetic crystal.

The two-soliton bound state solution we have obtained in this paper admits a transparent interpretation in each of the above physical 
situations.
In the
context of vibrated water troughs and chains of coupled 
pendula, the Bloch-N\'eel 
complex describes a patch oscillating $180^{\circ}$\ out of phase
with the rest of the chain or channel. A similar interpretation arises in optical parametric oscillators; there, a bound state of two dark
solitons
represents a localised fundamental field domain with a $180^{\circ}$\ phase difference from the rest of the cavity. In the context
of ferromagnetism, the stationary complex corresponds to 
a magnetic bubble, i.e. an ``island'' of one stable phase in the ``sea'' of the other
one.

The question of stability of the two-soliton bound 
state is beyond the scope of our current investigation.
The answer will obviously depend on whether Eq.(\ref{ODE}) is considered as a stationary reduction of the Ginsburg-Landau,
Klein-Gordon or nonlinear Schr\"odinger equation [i.e. Eqs.(\ref{2}), (\ref{KG}) or (\ref{NLS})].  To illustrate the 
model-dependence of stability properties of one and the same 
solution, it is instructive to bring up the example of a free-standing
N\'eel wall, Eq.(\ref{Neel}). Let, for example, $h < \frac 13$. 
If the N\'eel wall is considered as a stationary solution of
the Ginsburg-Landau equation (\ref{2})
or of the Klein-Gordon equation (\ref{KG}),
then it is found to be unstable while the Bloch wall  is stable \cite{Coullet1990,Skryabin,SubTrull,Ivanov}. 
On the contrary, {\it both} Bloch and N\'eel walls are stable \cite{Ourpaper} when considered as 
stationary solutions of the parametrically driven NLS equation, Eq.(\ref{NLS}).  
We are planning to return to the issue of stability of the bound states in future publications.

Finally, it is appropriate to mention that Eq.(\ref{ODE}) appeared in one more optical context, namely that of birefringent optical fibers.
The vector nonlinear Schr\"odinger equation for pulses travelling in a birefringent fiber was derived by Menyuk \cite{Menyuk}:
\begin{subequations}
\label{biref3}
\begin{eqnarray}
i\left(\frac{\partial E_1}{\partial t} + \delta \frac{\partial E_1}{\partial x} \right)+ \frac 12
\frac{\partial^2 E_1}{\partial x^2} - \left( |E_1|^2 + \frac 23 |E_2|^2\right)E_1  \nonumber \\ 
- \frac 13 E_2^2 E_1^* e^{-4iht} =0, \quad \quad \label{Biref3a} \\
i\left(\frac{\partial E_2}{\partial t}  -\delta \frac{\partial E_2}{\partial x} \right)+ \frac 12
\frac{\partial^2 E_2}{\partial x^2} - \left(\frac 23 |E_1|^2 + |E_2|^2 \right)E_2 
\nonumber \\ - \frac 13 E_1^2 E_2^* e^{4iht} =0. \quad \quad \label{Biref3b} 
\end{eqnarray}
\end{subequations}
Here $\delta$\ is proportional to the difference of group velocities of the fast and slow linearly polarised modes (whose envelopes are
described by $E_1$\ and $E_2$), and $h$\ measures the mismatch of the corresponding propagation constants. Equations
(\ref{Biref3a})-(\ref{Biref3b}) are written in a frame moving with the average of the group velocities; the choice of coefficients
corresponds to the fiber in the regime of normal dispersion. If one assumes that the difference of the two group velocities is so small that
it can
be neglected, while the difference of the propagation constants is nonnegligible (though possibly small), then the substitution 
\[ E_1 = U  e^{-i(h+1)t}, \quad E_2 = V e^{i(h-1)t},\]
takes Eq.(\ref{biref3}) to 
\begin{subequations}
\label{biref2}
\begin{eqnarray}
i\frac{\partial U}{\partial t}  + \frac 12
\frac{\partial^2 U}{\partial x^2} - \left( |U|^2 + \frac 23 |V|^2\right)U  \nonumber \\ 
- \frac 13 V^2 U^* + (1+h)U =0, \label{Biref2a} \\
i\frac{\partial V}{\partial t}  + \frac 12
\frac{\partial^2 V}{\partial x^2} - \left(\frac 23 |U|^2 + |V|^2 \right)V
\nonumber \\ - \frac 13 U^2 V^* + (1-h)V =0.  \label{Biref2b} 
\end{eqnarray}
\end{subequations}
The above assumption ($\delta = 0$, $h \neq 0$) can be justified in the case of the {\it bright} 
solitons, i.e. in the anomalous dispersion regime.
In that case, Blow, Doran and Wood demonstrated the existence of bound pairs of bright solitons where each soliton is polarised along a
different birefringence axis \cite{BDW}. Later, these solutions were explicitly constructed by Tratnik and Sipe \cite{TS}. Whether the
assumption $\delta = 0$, $h \neq 0$\ can be justified in the case of the {\it dark}
solitons is an open question which is beyond the scope of our
investigation. Here, we simply note that for time-independent real fields $U$\ and $V$, 
Eqs.(\ref{biref2}) reduce to our Eq.(\ref{ODE}) with  $\psi = U+iV$. 
 This fact was used by Christodoulides \cite{Ch} who obtained the Bloch and N\'eel walls for the weakly
birefringent fiber (under the above assumption). Our solution (\ref{H36})  is a bound state of the ``dark'' and ``bright-dark''
vector solitons of
Christodoulides.

\acknowledgments
IB was supported by the NRF of South Africa under grant 2053723, by Johnson Bequest Fund and the URC of the University of Cape Town. 
SW was supported by the NRF
of South Africa and by the URC of the University of Cape Town.

\end{document}